# Fabrication of Flexible Oriented Magnetic Thin Films with Large in-plane Uniaxial Anisotropy by Roll-to-roll Nanoimprint Lithography.

Rukshan M. Thantirige[1], Jacob John[2], Nihar R. Pradhan[3], Kenneth R Carter[2], Mark T. Tuominen[1]

[1] Department of Physics, University of Massachusetts, 666 N. Pleasant St, Amherst, MA, 01003, USA

[2] Department of Polymer Science and Engineering, University of Massachusetts, 120 Governors Dr, Amherst, MA, 01003, USA

[3] National High Magnetic Field Laboratory, 1800 E. Paul Dirac Drive, Tallahassee, FL-32310, USA

**Abstract:** Here, we report wafer scale fabrication of densely packed Fe nanostripe-based magnetic thin films on a flexible substrate and their magnetic anisotropy properties. We find that Fe nanostripes exhibit large in-plane uniaxial anisotropy and nearly square hysteresis loops with energy products $(BH)_{max}$ exceeding 3 MGOe at room temperature. High density Fe nanostripes were fabricated on 70 nm flexible polyethylene terephthalate (PET) gratings, which were made by roll-to-roll (R2R) UV nanoimprintlithography technique. Observed large in-plane uniaxial anisotropies along the long dimension of nanostripes are attributed to the shape. Temperature dependent hysteresis measurements confirm that the magnetization reversal is driven by non-coherent rotation reversal processes.

## I. INTRODUCTION

Thin films based on arrays of densely packed nanostructures are of great interest in novel applications and fundamental studies as they exhibit unique magnetic and electrical[1-3] properties with greater designed controllability than those of their bulk counterparts. As an example, thin films of magnetic nanostripes are widely used in studying domain wall dynamics induced by spin polarized currents and magnetic fields[4-6] due to their potential use in applications in information storage and logic devices[7-10], cell biology[11] and more recently in manipulating superconductivity[12]. In addition, a wide range of applications where mechanical flexibility is essential (flexible electronics), such as flexible solar cells, electronic paper, biomedical devices and sensors for non-rigid and non-planer surface detection[13-18] demand fabrication of these nanostructures and devices on flexible substrates. This study focuses on fabrication of magnetic nanostripes on a flexible substrate by roll-to-roll imprinting technology for magnetic thin film based devices with greater mechanical flexibility.

Previous research on fabrication of nanostripe samples have been achieved by utilizing both top down and bottom up techniques[19-22]. Although bottom up methods such as epitaxial growth can produce nanostructures with ultrasmall dimensions compared to traditional lithography based methods, their usage is limited due to requirement of ultrahigh vacuum conditions and restrictive lattice match between substrates and materials. In addition, such epitaxially grown samples show room temperature superparamagnetism[23,24] due to size effects, and are sensitive to defects. In contrast, template based methods have been explored and remain popular for their high precision and great designed controllability[20,25], although such techniques usually results nanostructures with larger dimensions. Shallow angle deposition of materials onto pre-patterned or vicinal templates is one common route exploited by many researchers[21,26,27]. In this method, materials are deposited at smaller angles ($< 4^0$) such that the deposition flux directed towards one side of the terrace while the other side being masked. Arora *et. al.*[28] followed this technique to fabricate Co nanostripe thin films which exhibit room temperature ferromagnetism and large in-plane coercivities up to 920 Oe.

In this study, we fabricated high quality wafer scale Fe-nanostripe thin films by deposition of materials through e-beam evaporation onto patterned topographical gratings on a substrate made by UV-assisted nanoimprint lithography (UV-NIL)[29-31]. We chose polyethylene terephthalate (PET) substrate due to its ready availability, flexibility, high mechanical, chemical stability and low cost. These properties make PET an attractive candidate for mass production by direct patterning with roll-to-roll nanoimprint lithography for high-throughput flexible device fabrication[32,33]. Fe nanostripes fabricated in this study exhibit nearly square hysteresis and $(BH)_{max}$ up to 3 MGOe along the long dimension of the nanostripes at room temperature, in contrast to samples made on planar PET substrates.

## II. EXPERIMENTAL PROCEDURES

The 70 nm wide, 50 nm deep topographical gratings were fabricated on PET films using roll-to-roll UV nanoimprint lithography, employing a roll-to-roll nanoimprinter fitted with *perfluoropolyether acrylate* (PFPE) based molds. The UV photoresist employed was Norland 81 (Norland Inc., USA) and it was used as received. For proof-of-concept purposes, gratings were diced into 3 mm x 3 mm pieces prior to magnetic material deposition (although continuous roll-to-roll deposition is also feasible). A detailed description of the roll-to-roll NIL process can be found elsewhere[34]. Fe thin films of 5 nm to 45 nm thickness were deposited by electron beam evaporation on PET gratings at normal incidence in high vacuum of $5\times10^{-7}$ torr. A 3 nm layer of Ag was also deposited on Fe as a capping layer to protect Fe from oxidation. Deposition rates were kept constant at 0.05 nm/s for all samples to promote continuous film growth. Films were also deposited on planar PET pieces with the same dimensions under identical deposition conditions as a control. Figure 1 illustrates the key steps of fabrication by roll-to-roll nanoimprinting process.

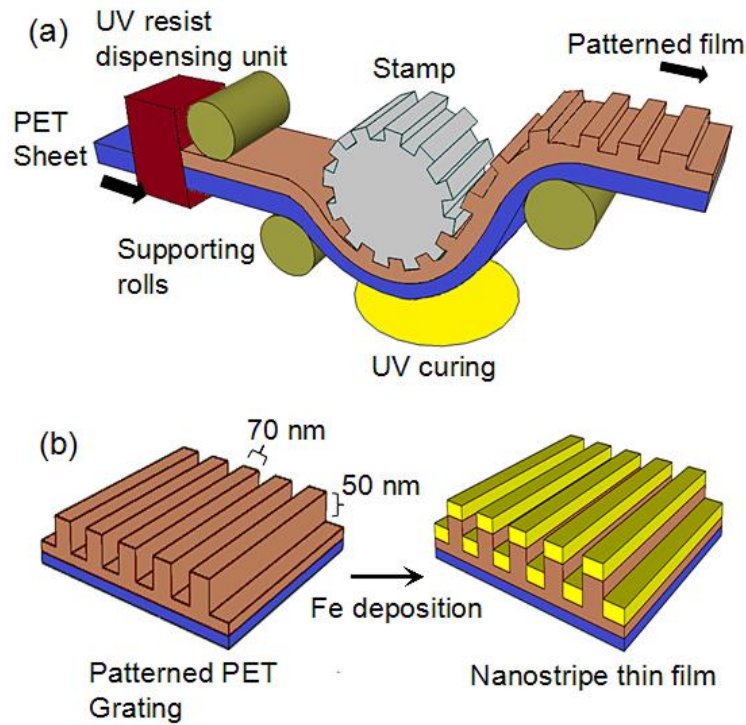

Figure 1. (a) Schematic representation of the UV-assisted roll-to-roll nanoimprint lithography process used in this work, and (b) fabrication of nanostripe-based thin film by metal evaporation.

Topography and morphology of each sample were analyzed by Atomic Force Microscopy (AFM)

and Scanning Electron Microscopy (SEM) techniques. In-plane magnetic properties were measured by SQUID magnetometry (Quantum Design MPMS-7T) and measurements were taken along ($H_{\parallel}$) and across ($H_{\perp}$) the long dimension of the nanostripes at various temperatures from 300 K to 10 K.

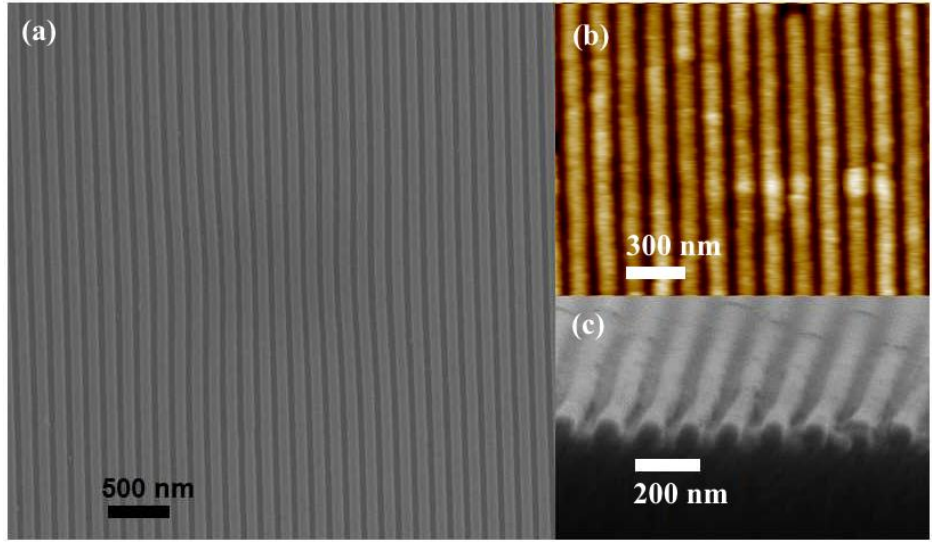

Figure 2. Fe nanostripes formation on PET grating upon e-beam evaporation of Fe at normal incidence. (a) SEM (JOEL 7001F) micrograph of 15 nm thick nanostripes. (b) AFM (Veeco NanoScope IV) micrograph of 15 nm thick nanostripes, shows degree of defects and discontinuity, and (c) Cross section SEM micrograph of 15 nm thick Fe nanostripe thin film. This shows the material deposition on sidewalls. Scale bar of the figure (a), (b), (c) are 500 nm, 300 nm and 200 nm, respectively.

## III. RESULTS AND DISCUSSION

We first present structural and morphological characterization obtained for a 15 nm thick nanostripe sample. Figure 2 depicts the typical assembly of Fe nanostripes on PET gratings. Figure 2(a), top-view SEM image, shows that these nanostripes are highly periodic and have high degree of continuity with low defect concentration. However, the average width of nanostripes deposited on peaks of the grating is 85 nm, hence appear to be slightly wider than the grating width of 70 nm. Figure 2(b), a representative AFM image, reveals additional information such as the roughness and imperfections along each nanostripe, which may have been transferred from the substrate or a result of dewetting between Fe and PET. The analysis of AFM images reveals that the average width of nanostripes is 79 nm and the step height has a low statistical variation of +/- 2 nm (peak-to-peak). The cross section SEM image shown in Figure 2(c) confirms the material deposition on sidewalls that creates thinner nanostripes on sidewalls that can induce pinning effect on 70 nm wide principal nanostripes.

To understand the effect of high aspect ratio (length/width), we performed in-plane magnetic measurements along ($H_\parallel$) and across ($H\perp$) nanostripes. Figure 3(a) and (b) show the typical hysteresis curves measured for 27.5 nm thick nanostripes at 300 K and a representative low temperature of 50 K, respectively. The applied field was varied in the range of +/- 20 KOe, but only +/- 2 KOe regime is shown for clarity. At both 300 K and 50 K, $H_\parallel$ reaches faster towards saturation, and their $M_r/M_S$ ratios exceed 90% as a result of shape dominated anisotropy. Also, it can be noticed that the $H_C$ rises with decreasing the temperature from 185 Oe at 300 K to 260 Oe at 50 K for 27.5 nm thick nanostripes. This observation of enhancement of $H_C$ at lower temperatures is in line with general expectation of decrease in thermal fluctuations at lower temperatures. Further, $H_\parallel$ hysteresis curves have 'shoulders' at both 300 K and 50 K as shown in Figure 3(a) and 3(b). This two-step reversal can be understood as a result of pinning induced by thinner nanostripes formed on sidewalls that have larger aspect ratios (high $H_C$), on principle nanostripes at edges.

Figure 4 shows in-plane demagnetization curves of nanostripes with various thicknesses from 5

nm to 45 nm and a representative planar sample of 15 nm thickness at 300 K, and elucidates the role of thickness on anisotropy and reversal. Measurements of 20 nm, 27.5 nm and 35 nm thick nanostripes are not shown for clarity. As the figure depicts, we can see that there is a clear correlation between the nanostripe thickness and magnetic properties. Also, a direct comparison between 15 nm planar sample and 15 nm nanostripes explicates how the shape anisotropy enhances hard magnetic properties. The thickness dependence can further be elaborated with calculated and extracted data from Figure 4, shown in Table 1. If we consider $H_C$, it first increases with the nanostripe thickness from 194 Oe at 5 nm to reach the maximum of 257 Oe at 15 nm, and gradually decreases to 162 Oe at 45 nm. The $(BH)_{max}$ and reduced remanence ($M_r/M_S$) also follow the same trend giving a maximum of 3.5 MGOe and 0.96 for the 15 nm thick nanostripes, respectively. This increase in $H_C$ with the thickness for thin nanostripes is in-line with previous reports[28,35,36] and also agrees with mean field studies that predict a linear increase with the thickness[37,38] In addition, at low thicknesses the in-plane orientation of spins can be challenged by the substrate roughness and de-wetting behavior that promotes the perpendicular orientation, which lowers the coercivity further. In order to interpret the decrease in $H_C$ for thicker nanostripes (>15 nm), we need to consider the influence of the reversal mode, as large single domain magnets are more curling-dominated that lowers the demagnetizing field, hence the shape anisotropy. In addition, magnetic dipolar interactions that grow with nanostripe thickness reduce the effective $H_C$, as such interactions prefer anti-parallel arrangement of magnetic spins. As an example, if $N$ nanostripes are reversed by dipolar interactions, assuming each reversal reduces the total magnetostatic energy by $E_V$, the interaction energy between two nanostripes, the effective $H_C$ of the sample can be given by[28, 39],

$$H_C = \frac{2K}{\mu_0 M_S}\left[1 - \left(\frac{N|E_V|}{K}\right)^{\frac{1}{2}}\right] \qquad (1)$$

Where the prefactor $2K/\mu_0 M_S$ denotes the intrinsic coercivity due to anisotropy $K$, of an isolated nanostripe.

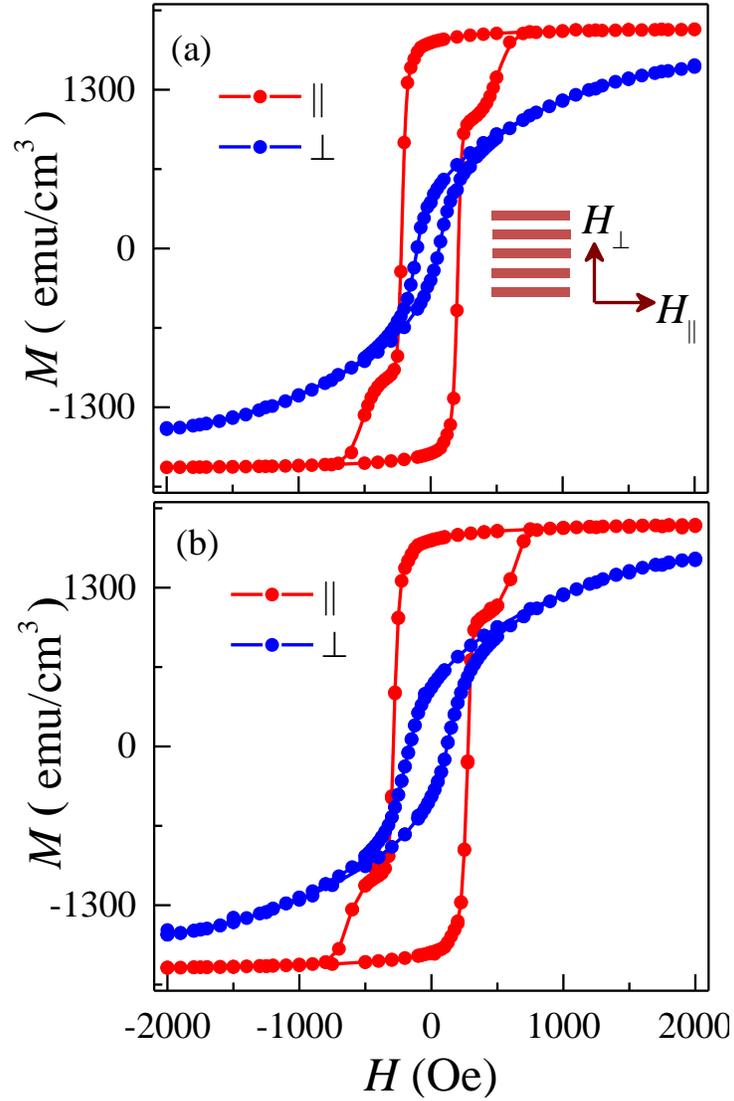

Figure 3. *M-H* curve of 27.5 nm thick Fe nanostripes with field applied in-plane along ($H_{\parallel}$) and across ($H\perp$) the long-axis of stripes at (a) 300 K and (b) 50 K. The shoulders in $H_{\parallel}$ curves are due to pinning effect induced by sidewalls at edges of principle nanostripes.

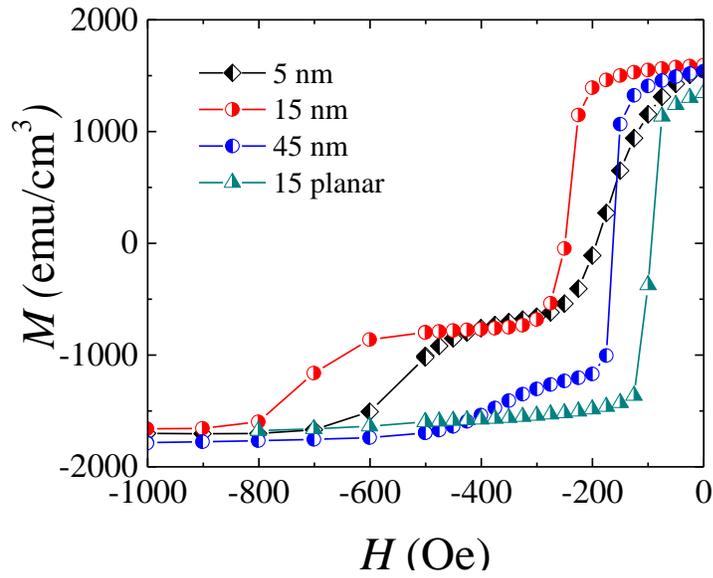

Figure 4. Demagnetization curves of 5, 15, 45 nm thick nanostripes films and 15nm thick planner film at 300K for field applied in-plane along nanostripes. Nanostripes have wider hysteresis in-contrast to planar sample and the $H_C$ of nanostripes has a clear thickness dependence.

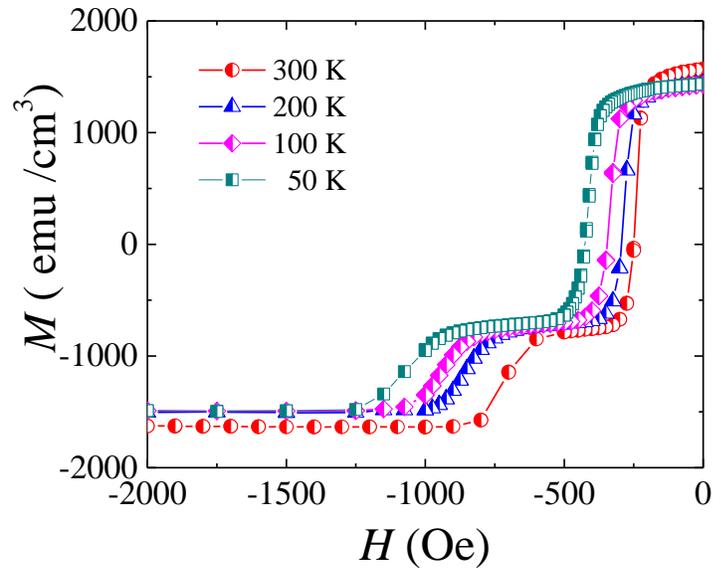

Figure 5. Demagnetization curves of 15 nm thick nanostripes at 300K, 200 K, 100 K and 50 K for field applied in-plane along nanostripes. Hysteresis widens with reducing the temperature due to low thermal fluctuations.

Table 1. Reduced remanence ($M_r/M_S$), coercivity ($H_c$) and maximum energy product $(BH)_{max}$ for Fe nanostripe samples with 5 - 45 nm thickness ($t$) (calculated/extracted from Figure 5).

| Sample | $t$ (nm) | $M_r/M_S$ | $H_C$ (Oe) | $BH_{max}$ (MGOe) |
|--------|----------|-----------|------------|-------------------|
| 1 | 5 | 0.78 | 194 | 1.5 |
| 2 | 10 | 0.89 | 215 | 2.2 |
| 3 | 15 | 0.96 | 257 | 3.5 |
| 4 | 20 | 0.96 | 247 | 3.1 |
| 5 | 27.5 | 0.92 | 185 | 2.7 |
| 6 | 35 | 0.87 | 177 | 2.3 |
| 7 | 45 | 0.86 | 162 | 2.1 |

As expected for large aspect ratio nanostripes, the hysteresis loop along the long axis is closer to a square and reduced remanence ($M_r/M_S$) is closer to 1 for all thicknesses except 5 nm sample, which is very sensitive to template roughness and imperfections. However, the observed switching field values fall well below the theoretical values given by $2\pi M_S$ (for Fe, $2\pi M_S = 10.8$ kOe), suggesting that the reversal is not governed by coherent reversal mode but by curling, buckling, domain wall motion or by any combination of them[40].

To understand the reversal mechanism, we performed hysteresis measurements at varies temperatures from 300 K to 10 K for selected samples. Figure 5 shows the demagnetization curves of a typical 15 nm thick nanostripe sample at 300 K, 200 K, 100 K and 50 K. It should be noted that measurements taken at 250 K, 150 K, and 10 K are not shown here for clarity. As the figure illustrates, the hysteresis widens, hence the $H_C$ increases with lowering the temperature due to low thermal fluctuations. Also, it can be seen that $M_r$ slightly decreases at lower temperatures. $H_C$ variation with temperatures for four different samples, extracted from temperature dependent hysteresis measurements (Figure 5), is shown in Figure 6. The data (points) were fitted with the model proposed by He *et. al.*[41] (dashed lines) for temperature dependence of coercivity $H_C(T)$ of shape anisotropy dominated soft ferromagnetic structures. This model is the first term of equation (2), which is an extension of early work by Neel[42] and Brown[43] to study magnetic reversal process[36,41]. However, it can be seen that He's model and experimental data do not match well in the low temperature regime (< 100 K), in contrast to previous reports[28,36]. As an example, $H_C$ values at 10 K are about 50% higher than model predicted values. In order to match with data in the low temperature regime, the original equation was extended with a second term as shown in equation (2), which predicts an exponential decay with the temperature. So the extended equation can be given as,

$$H_C(T) = H_1(0) \frac{M_S(T)}{M_S(0)} \left[ 1 - \left( \frac{25 K_B T M_S^2(0)}{E_0 M_S^2(T)} \right)^{\frac{1}{\alpha}} \right] + H_2(0) \cdot \exp\left(-\frac{T}{T'}\right) \quad (2)$$

Here, $H_1(0)$ and $E_0$ are the coercivity at 0 K and the energy barrier of reversal governed by shape, respectively as predicted by He's model. $M_S(0)$ and $M_S(T)$ are the magnetizations at 0 K and T temperatures, respectively. The exponent α depends on the specific reversal mode, with α = 3/2 and α = 2 corresponding to curling mode and coherent rotation mode, respectively. Since the width of nanostripe is beyond the critical size for coherent rotation given by $2.08 A^{1/2} / M_S$ which is 12 nm for Fe[36], we fitted experimental data with α = 3/2 to estimate $H_1(0)$, $H_2(0)$, $E_0$ and T' (Table 2). Here, the temperature variation of saturation magnetization has been ignored as it is negligible for the temperature range in concern. Further, we found that $H_2(0)$ has a rough linear relation to the film thickness, which takes the form $H_2(0) = 410 - 8.7t$. The coercivity at 0 K,

$H_C(0)$ is the addition of $H_1(0)$ and $H_2(0)$ from which the activation volume $V^*$, the region that the reversal process is localized (nucleation core), can be calculated using the relation $E_0 = V^* M_S(0) H_C(0)$. Table 3 shows estimated values of $V^*$ and nucleation core size, $L$ ($V=L^3$) for selected samples. The size of the nucleation core (15-22 nm) is much smaller than the physical size of the nanostripe, confirming that the nanostripe as a whole does not undergo coherent reversal.

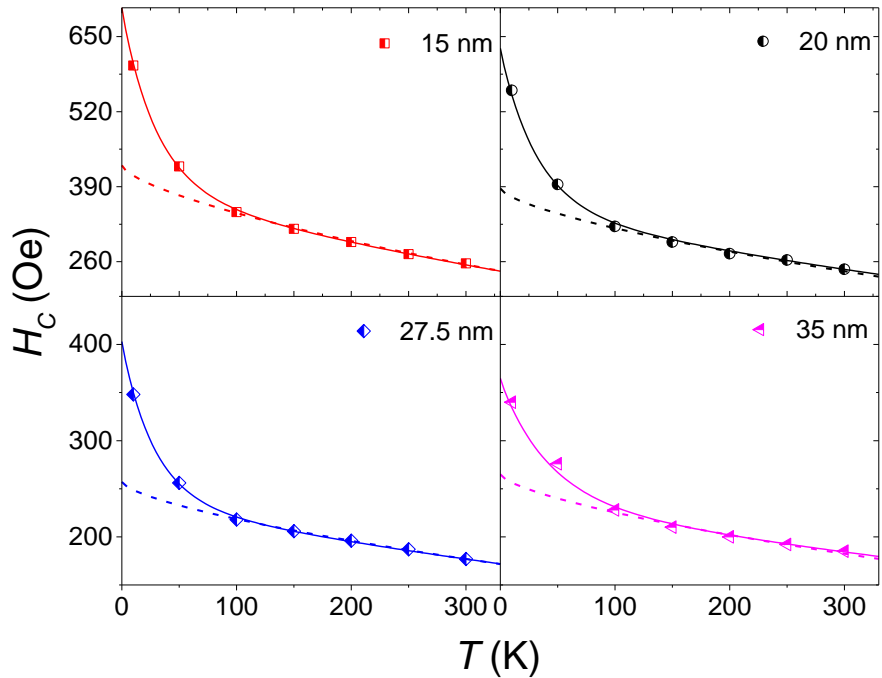

Figure 6. Temperature dependence of $H_C$ for selected samples with field applied parallel to nanostripes. The dashed and continues lines represents the fitted curve excluding and including the exponential decay term, respectively of equation (2) for $\alpha = 3/2$, respectively.

Table 2. Parameters of $H_1(0)$, $H_2(0)$, $E_0$ and T' estimated by fitting equation (2) with $H_C$ Vs. T by taking $\alpha = 3/2$ for selected samples.

| Sample | t (nm) | $H_1(0)$ (Oe) | $H_2(0)$ (Oe) | $E_0$ (eV) | T' (K) |
|---|---|---|---|---|---|
| 3 | 15 | 420 | 280 | 2.6 | 30 |
| 4 | 20 | 380 | 250 | 3.1 | 32 |
| 5 | 27.5 | 253 | 150 | 3.9 | 28 |
| 6 | 35 | 255 | 110 | 4.4 | 42 |

Table 3. Zero temperature coercivity $H_C(0)$, activation volume $V^*$ and nucleation core size $L$, estimated from parameters in Table 2.

| Sample | t (nm) | $H_C(0)$ (Oe) | $V^*$ (nm$^3$) | L (nm) |
|---|---|---|---|---|
| 3 | 15 | 700 | 3410 | 15 |
| 4 | 20 | 630 | 4518 | 16.5 |
| 5 | 27.5 | 403 | 8886 | 20.7 |
| 6 | 35 | 365 | 11069 | 22.3 |

## IV. CONCLUSIONS

Fe nanostripe based thin films exhibiting larger uniaxial anisotropies were fabricated by UV-assisted nanoimprint lithography. PET film was chosen as the substrate due to its good mechanical properties, low cost, and demonstrated high volume direct patterning capabilities in roll-to-roll nanoimprinting. The in-plane $H_C$ along nanostripes, induced by shape, changes with the film thickness giving the maximum value of 257 Oe for 15 nm thick nanostripes at room temperature. This variation of $H_C$ with film thickness has been attributed to the growing dipolar interaction with the material thickness. By combining magnetization measurements at different temperatures and extending the model predicted by He et. al.[41], we found that the magnetization reversal process is driven by non-coherent rotation reversal processes, and the size of the nucleation core is much smaller than the physical volume of the nanostripe. We propose that the 'shoulders' observed in easy axis hysteresis loops are due to pinning effect induced by narrow nanostripes formed on sidewalls, on edges of principle nanostripes. One noteworthy advantage of nanostripe based thin films fabricated in this work is that they are highly anisotropic but maintain the same magnetic moment per unit area as planar thin films, which can be further enhanced by reducing the width of nanostripes to a certain extent. To conclude, we believe that this straightforward fabrication method can be implemented for high volume fabrication of a range of future ferromagnetic nanoscale thin film based devices with great mechanical flexibility, where low cost and high performance will dominate future needs.

**ACKNOWLEDGMENTS**

This work was supported by NSF grants DMR-1208042 and NSEC Center for Hierarchical Manufacturing-CMMI-1025020. The authors wish to acknowledge John Nicholson and Ashan Vitharana for their technical advice and support.